\def\year{2022}\relax
\documentclass[letterpaper]{article} % DO NOT CHANGE THIS
\usepackage{aaai22}  % DO NOT CHANGE THIS
\usepackage{times}  % DO NOT CHANGE THIS
\usepackage{helvet}  % DO NOT CHANGE THIS
\usepackage{courier}  % DO NOT CHANGE THIS
\usepackage[hyphens]{url}  % DO NOT CHANGE THIS
\usepackage{graphicx} % DO NOT CHANGE THIS
\usepackage{gensymb}
\usepackage{booktabs}
\usepackage{caption} % DO NOT CHANGE THIS AND DO NOT ADD ANY OPTIONS TO IT
\DeclareCaptionStyle{ruled}{labelfont=normalfont,labelsep=colon,strut=off} % DO NOT CHANGE THIS
\usepackage{algorithm}
\usepackage{algorithmic}

\usepackage{newfloat}
\usepackage{listings}
\floatstyle{ruled}
\newfloat{listing}{tb}{lst}{}
\floatname{listing}{Listing}

\title{Contrastive Learning for Climate Model Bias Correction and Super-Resolution}
\author{
    Tristan Ballard,
    Gopal Erinjippurath
}
\affiliations{
    Sust Global\\
    San Francisco, California\\
}

\begin{document}

\maketitle

\begin{abstract}
Climate models often require post-processing in order to make accurate estimates of local climate risk. The most common post-processing applied is bias-correction and spatial resolution enhancement. However, the statistical methods typically used for this not only are incapable of capturing multivariate spatial correlation information but are also reliant on rich observational data often not available outside of developed countries, limiting their potential. Here we propose an alternative approach to this challenge based on a combination of image super resolution (SR) and contrastive learning generative adversarial networks (GANs). We benchmark performance against NASA’s flagship post-processed CMIP6 climate model product, NEX-GDDP. We find that our model successfully reaches a spatial resolution double that of NASA’s product while also achieving comparable or improved levels of bias correction in both daily precipitation and temperature. The resulting higher fidelity simulations of present and forward-looking climate can enable more local, accurate models of hazards like flooding, drought, and heatwaves.
\end{abstract}

\section{Introduction}

Global climate models by design are imperfect simulations of the physical world. While leading climate models like those in the Coupled Model Intercomparison Project phase 6 (CMIP6) incorporate known phenomena like the laws of thermodynamics, other phenomena like cloud condensation have no known equations and require developers to include imprecise estimates. What’s more, climate models are run at spatial resolutions too coarse to simulate key phenomena like convective precipitation, tropical cyclone dynamics, and local effects from topography and land cover (microclimates). This leads to a variety of known and unknown errors, or biases, in projections of fundamental variables like temperature and precipitation. 

Climate model errors reduce the accuracy of projections of climate hazards like heatwaves and flooding, motivating the development of bias correction methods. These methods (Section 2) generally involve deriving correction factors to better align modeled historical values with observed historical values. The correction factors are then applied to forward-looking modeled values and are widely implemented in the climate impacts community. Indeed, forward-looking estimates of future flood risk typically use bias-corrected precipitation rather than the raw climate model data \cite{Salman2018}. To enable local, accurate hazard models requires high fidelity, bias-corrected simulations of present-day and forward looking fundamental variables.

\begin{figure}[] % t = top
\includegraphics[width=8.5cm]{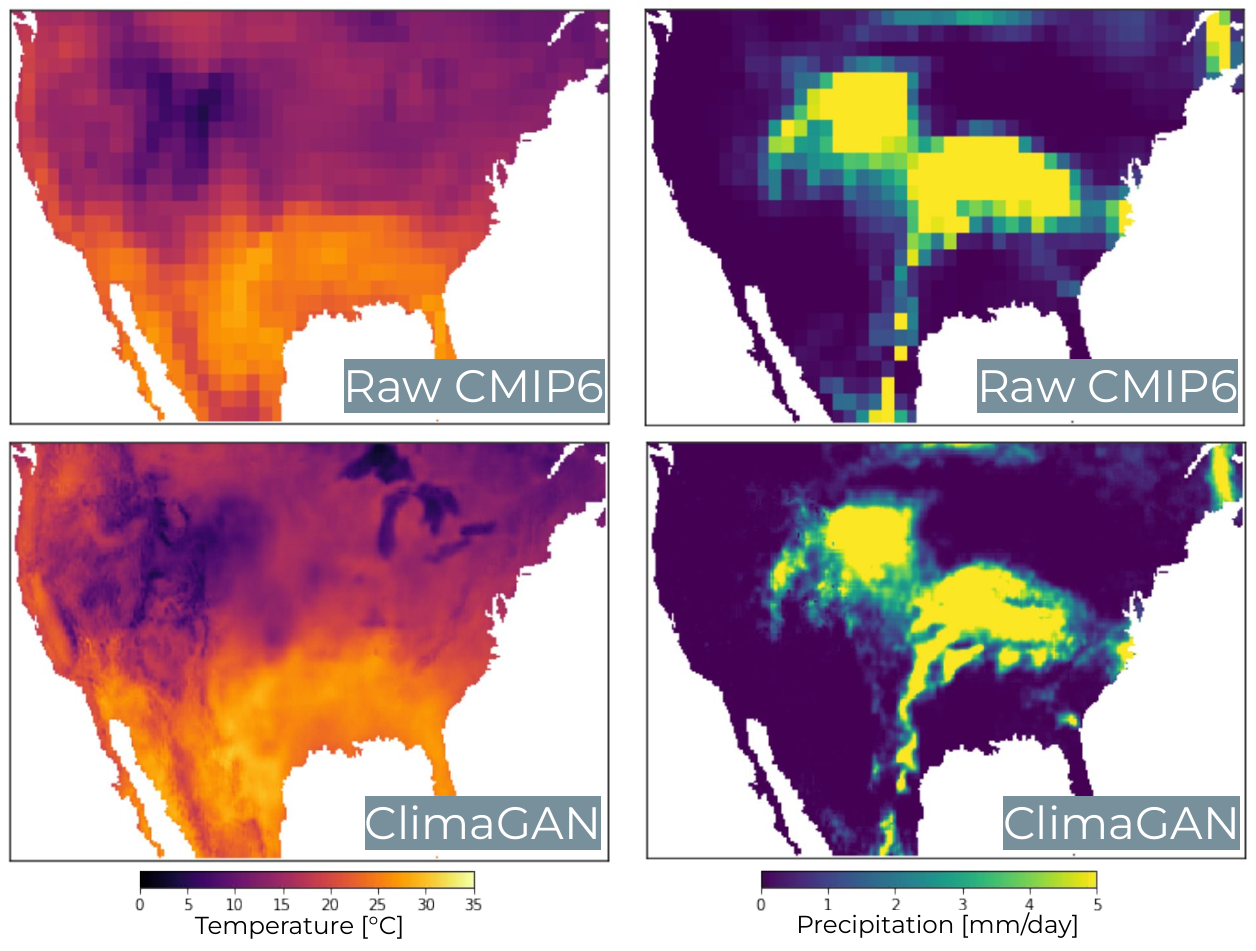}
\centering
\caption{Application of the ClimaGAN network to a CMIP6 test set image (May 18, 1994) yields bias-corrected and 4x (0.5\degree{} $\rightarrow$ 0.125\degree{}) super-resolution outputs. The network can be applied to CMIP6 daily simulations out to 2100.}
\label{fig1}
\end{figure}

Recent advances in AI including in image super-resolution (SR) and unpaired image-to-image translation suggest substantial promise to improve over existing bias correction methods. These AI models can flexibly incorporate multivariate and spatial relationships in ways not possible with existing approaches. For instance, AI-based SR has shown superior performance in enhancing the spatial resolution of wildfires \cite{Ballard2020}, precipitation \cite{cheng2020reslap, Vandal2017, Vaughan2021}, and wind \cite{Kurinchi2021, Stengel2020}. Meanwhile, unpaired generative adversarial networks (GANs) have shown promise in applications to temperature \cite{Fulton2021} and precipitation \cite{Fulton2021, Pan2021}.

Here we propose ClimaGAN, a novel SR and unpaired image-to-image translation GAN architecture operating on 3-channel geospatial datasets, incorporating temperature, precipitation, and elevation. We validate and compare ClimaGAN performance against a NASA benchmark algorithm, showcasing ClimaGAN performance on a leading CMIP6 model over a region spanning the contiguous U.S.

\section{Related Work}
There are several methods for bias-correcting and resolution enhancement (downscaling) of climate variables, but the predominant method implemented in the climate community is the bias-correction spatial disaggregation (BCSD) algorithm. For example, BCSD, proposed in 2002 \cite{Wood2002}, is the method used for NASA's flagship CMIP6 bias-corrected product \cite{Thrasher2022}(Section 3.4). The bias-correction portion of BCSD is achieved through simple quantile mapping between modeled and observed cumulative distribution functions. The resolution enhancement is achieved through application of Fourier transforms.  

The chief limitation of the (BCSD) algorithm used in NASA’s NEX-GDDP product \cite{Thrasher2022} is that it is a simple statistical method incapable of  incorporating auxiliary datasets or spatial variability. For example, the only data that can be used to bias correct a modeled temperature dataset using BCSD is an observed temperature dataset. However, we know that temperature biases are tightly linked to local features like elevation \cite{Lun2021}. BCSD also implements bias correction independently for each pixel, ignoring spatial correlation structure that can provide useful signal for further reducing biases. 

BCSD also does not permit multivariate relationships between climate variables, despite the fact that most climate variables covary. Bias correcting temperature independently from precipitation, for example, can inadvertently introduce unrealistic relationships, particularly for extremes \cite{Li2014}. Bivariate BCSD has been proposed but has not been widely adopted \cite{Li2014}.

We are aware of two recent AI-based approaches for climate model bias correction, but neither incorporate spatial resolution enhancement. Both approaches are based on unpaired image-to-image translation, with one adapting the cycleGAN framework \cite{Pan2021} and the other using UNIT \cite{Fulton2021}. Extreme learning machines have also been proposed as an alternative to BCSD \cite{Zhang2017}.

\section{Data}
Geospatial data coded as input images to the model architecture (Fig. 2) have 3 channels, corresponding to maps of daily temperature, daily precipitation, and elevation. Low resolution (LR) input images come from CMIP6 climate model simulations regridded to a common 0.5° resolution, while high resolution (HR) input images come from observed weather data regridded to a common 0.125° (14km) resolution. 

The study area covers the contiguous U.S., southern Canada, and northern Mexico, spanning 23°N and 49°N and 125°W and 65°W (Fig. 1). The LR input images are of dimension 54x120x3 while HR images are of dimension 216x480x3.  

We train the model on 24 years of data from 1985 to 2014, setting aside 6 years [1990, 1994, 2000, 2004, 2008, 2012] of data in that period for testing. This results in 8,756 daily images for training and 2,194 daily images for testing.

\subsection{CMIP6 climate model simulations}
We demonstrate the ClimaGAN network with the U.S. National Oceanic and Atmospheric Administration's Geophysical Fluid Dynamics Laboratory model GFDL-CM4, a leading CMIP6 model \cite{Held2019}. The ClimaGAN network can be retrained and applied to any of the CMIP6 models, an extension of the current research we are actively pursuing (Section 7). The network can also be applied to corresponding CMIP6 forward looking (2015-2100) projections to derive estimates of future hazards (not shown). 

While CMIP6 models simulate a range of climate variables, we focus here on simulations of daily maximum temperature and daily precipitation because these are often needed to derive climate hazards. The historical CMIP6 simulations incorporate known values of carbon emissions, solar activity, and volcanic eruptions, among other inputs. 

\subsection{Elevation}
We incorporate elevation data from the National Center for Atmospheric Research \cite{TerrainBase} as a supplementary feature to inform bias correction. Elevation is an important driver of local climate, so we expect it to be informative in bias correcting both temperature and precipitation.

\subsection{Observations}
We use the European Centre for Medium-Range Weather Forecasts (ECMWF) ERA5-Land data for observed daily maximum temperature and daily precipitation \cite{Munoz2021}. The reanalysis data has global coverage at approximately 9km resolution over land, which we regrid to a coarser 0.125° for the analysis.

\subsection{NASA NEX-GDDP benchmark product}
We benchmark model performance against NASA's flagship CMIP6 bias-corrected product, NEX-GDDP \cite{Thrasher2022}. NEX-GDDP is based on the BCSD algorithm (Section 2). We use the daily maximum temperature and precipitation NEX-GDDP data corresponding to the same GFDL-CM4 model, such that outputs between the two methods are directly comparable. NEX-GDDP data is available at 0.25° resolution. 

Beyond the technical limitations of NEX-GDDP (Section 2), practical limitations for users are that it is not updated with the latest observational datasets and covers only a few variables and climate scenarios. The amount of available observational data is projected to increase substantially with the release of new satellite and sensor datasets, yet NEX-GDDP will not begin to incorporate that new data until the release of CMIP7 years from now, if at all. This means that any advances in monitoring in data-poor regions, such as in many developing countries, will not be incorporated. Further, NEX-GDDP only covers 9 climate variables from the ScenarioMIP project, despite there being hundreds of other variables and MIP projects within CMIP6 of interest to researchers, limiting its scope.

\begin{figure*}[t] % t = top
\includegraphics[width=0.9\textwidth]{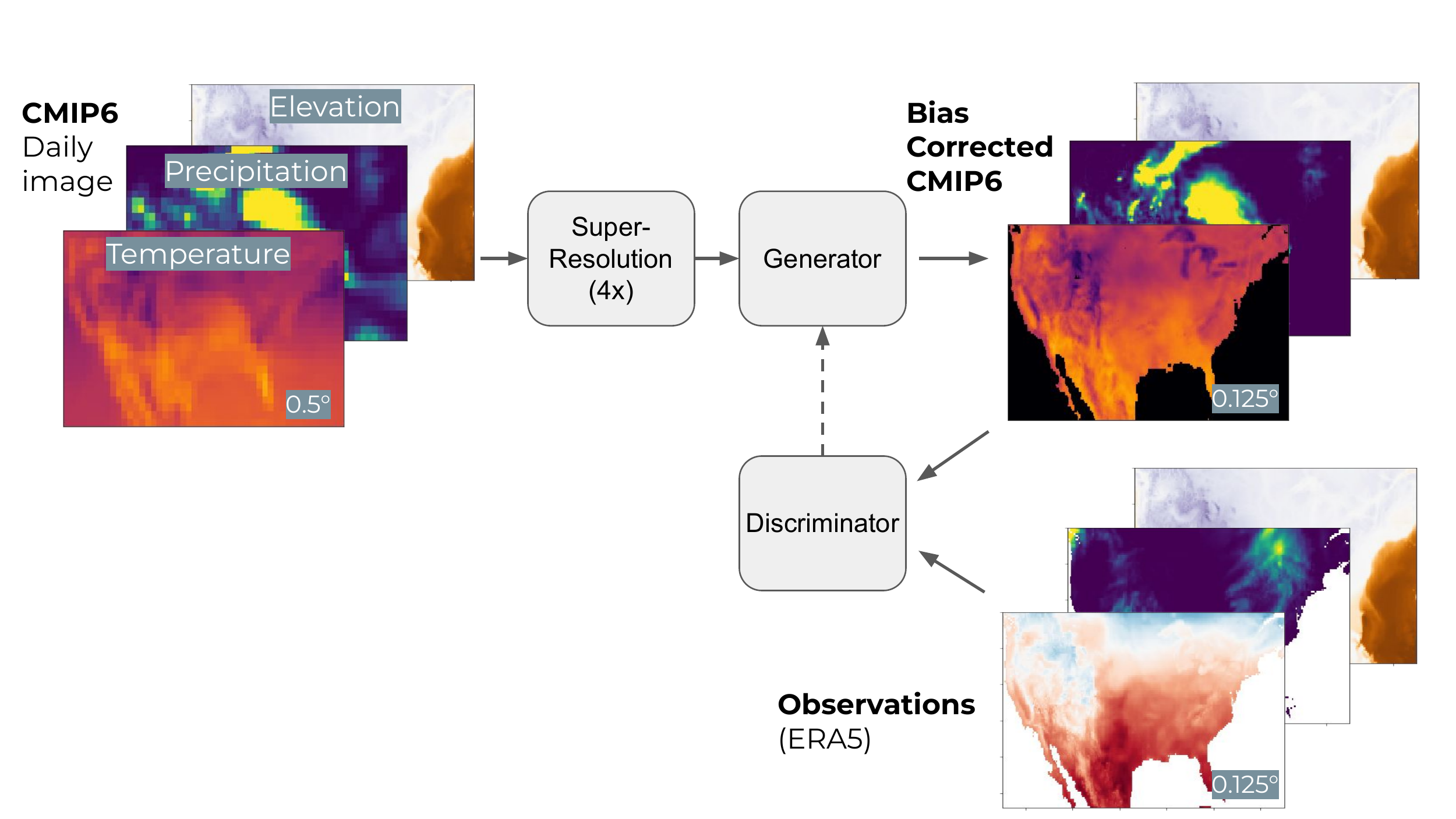}
\centering
\caption{The ClimaGAN network takes as input daily CMIP6 climate data, as well as supplementary features like elevation, and outputs corresponding high-resolution, bias-corrected daily data. The network combines two key modules: super-resolution (SR) and a contrastive unpaired translation GAN. The SR layers enhance spatial resolution by 4x (0.5\degree{} $\rightarrow$ 0.125\degree{}), while the GAN iteratively learns to bias-correct climate model inputs from comparisons with real-world observations.}
\label{fig3}
\end{figure*}

\section{Methodology}
\subsection{ClimaGAN Architecture}
We identified four key design goals for our network architecture:
\begin{itemize}
  \item Unpaired image-to-image translation
  \item Content preservation
  \item Spatial resolution enhancement (super-resolution)
  \item Multivariate input and output variables
\end{itemize}

Unpaired image-to-image translation is required because the daily output from a CMIP6 model is not expected to directly match observations for the corresponding date, a challenge for typical bias-correction methods. For example, CMIP6 temperature simulations for Jan 1, 2010 are not, by design, expected to match observed conditions on that date. They instead are expected to provide a realistic simulation of what the weather could have been on that date.

Content preservation is the idea that the bias-corrected output variables should maintain the content of the CMIP6 inputs while taking on the appearance of real-world conditions. Content preservation in the context of GANs is typically preserved through adding a cycle-consistency loss term \cite{Zhu2017}.

\begin{figure*}[t] % t = top
\includegraphics[width=1.0\textwidth]{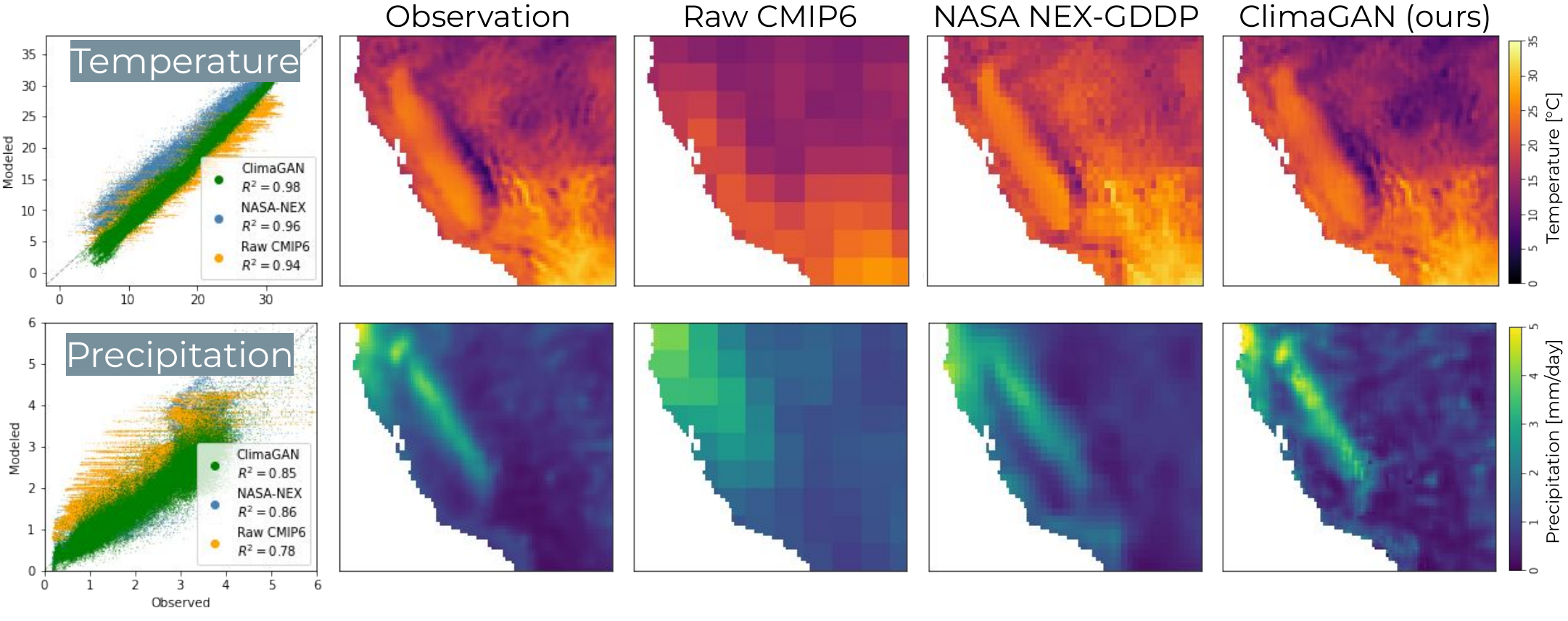}
\centering
\caption{ClimaGAN quantitatively and qualitatively enhances CMIP6 mean daily temperature (top) and daily precipitation (bottom) in a held out test set (\textit{n}=2,194 daily images). The scatter plots show the pixel-by-pixel correlations against observations across the U.S., while the maps show the southwestern U.S.}
\label{fig2}
\end{figure*}

To achieve these design goals, we designed a network (ClimaGAN) that combines super-resolution and a contrastive unpaired translation GAN (Fig. 2). The input LR images passed through the network first go through two SR layers that enhance spatial resolution by 4x. These SR images are then passed through a generator network. The discriminator compares the output images with observation images to determine which image is 'real' (observation) and which image is 'fake' (bias-corrected and super-resolved CMIP6). 

The generator and discriminator networks along with the super-resolution layers are trained concurrently. As the generator and discriminator improve, so does the level of bias-correction, creating output climate  images that are increasingly difficult to distinguish from real-world observations. The generator consists of 9 Resnet blocks in between two upscaling layers (‘encoder’) and two downscaling layers (‘decoder’). The discriminator consists of 3 convolutional layers. The network contains approximately 14M parameters, and we train it for 20 epochs on an NVIDIA Tesla A100 GPU. 

One of the key advances of this network is the implementation of a contrastive unpaired translation GAN. The contrastive unpaired translation is an advancement in GANs released in 2020 from the team who created cycleGAN, a leading framework for unpaired image-to-image translation \cite{Park2020}. Contrastive unpaired translation appears to outperform cycleGANs in both accuracy and efficiency \cite{Park2020}. Briefly, the network incorporates a InfoNCE loss term \cite{Oord2018} in addition to the adversarial loss of a standard GAN \cite{Park2020}. The InfoNCE loss works by sampling patches of the output image and ensuring that the samples are similar to the corresponding patches of the input image. At the same time, the InfoNCE loss discourages the sampled patches from being too similar to other patches of the input image. This loss term achieves content preservation. Further details on InfoNCE loss and the corresponding architecture additions can be found in Park \textit{et al}.\cite{Park2020} We use the same default hyperparameters for the number and size of patches as in Park \textit{et al}. \cite{Park2020}

\subsection{Validation}
To validate the model, we measure correspondence between observations and ClimaGAN output on a held out test set (Section 3). We selected four statistical measures to assess the fidelity of model simulations in representing the observed statistical distribution: mean, standard deviation, skew, and the 98\textsuperscript{th} percentile. The 98\textsuperscript{th} percentile reflects the ability of the model to capture extremes. These statistics are computed for each pixel and then plotted as maps (Fig. 3) or aggregated across pixels using R\textsuperscript{2} (Tables 1 and 2).

\section{Results}
We find that ClimaGAN substantially improves CMIP6 input simulations of daily temperature and precipitation, not only enhancing spatial resolution 4x to 0.125\degree{} but also leading to reductions in bias when evaluated on the held out test set. 

We first evaluate performance enhancement qualitatively by comparing maps of observed conditions against modeled (Fig. 3). Figure 3 shows mean conditions over the collection of daily test set images. Visually, the ClimaGAN-enhanced CMIP6 conditions much better match observed compared with the raw CMIP6 input, capturing local spatial variability with higher accuracy. In California, the Central Valley is reflected clearly in enhanced temperatures, while the eastward Sierra Nevada mountains are reflected by a band of elevated precipitation, distinctions not immediately apparent in the original CMIP6 data (Fig. 3).

\begin{table}[h]
\centering
\begin{tabular}{@{}llllll@{}}
\toprule
Temperature          & Mean     & SD       & Skew & Q98        \\ \midrule
ClimaGAN (ours)    & \textbf{0.98} & \textbf{0.97} & 0.26  & \textbf{0.94}      \\
NASA NEX-GDDP    & 0.96 &  0.90 &  \textbf{0.69}  &  0.86      \\ 
Raw CMIP6 & 0.94 & 0.88 & 0.42   & 0.75   \\ \bottomrule
\end{tabular}
\caption{ClimaGAN applied to daily maximum temperature shows an enhancement of raw CMIP6 inputs in out of sample test set years (\textit{n}=2,194 daily images) across all four evaluation metrics over the U.S. and outperforms NASA's product except for distribution skew. Q98 = 98\textsuperscript{th} percentile.}
\label{table1}
\end{table}

\begin{table}[h]
\centering
\begin{tabular}{@{}llllll@{}}
\toprule
Precipitation          & Mean     & SD       & Skew & Q98        \\ \midrule
ClimaGAN (ours)    & 0.85 & 0.80 & 0.39    & 0.78      \\
NASA NEX-GDDP    & \textbf{0.86} &  \textbf{0.81} &  \textbf{0.42}  &  \textbf{0.80}      \\ 
Raw CMIP6 & 0.78 & 0.69 & 0.37   & 0.72   \\ \bottomrule
\end{tabular}
\caption{ClimaGAN applied to daily precipitation shows an enhancement of raw CMIP6 inputs in out of sample test set years (\textit{n}=2,194 daily images) across all four evaluation metrics over the U.S., though NASA's product slightly outperforms ClimaGAN. Q98 = 98\textsuperscript{th} percentile.}

\label{table2}
\end{table}

Next, we evaluate performance enhancement quantitatively in the held out test set, finding ClimaGAN improves over raw CMIP6 data across all four statistical measures for precipitation and across three of four statistical measures for temperature (Tables 1 and 2). For temperature, mean daily temperature improves from an R\textsuperscript{2} of 94\% with the original CMIP6 data to an R\textsuperscript{2} of 98\% after applying ClimaGAN (Fig. 3; Table 1). We also find that extreme temperature, represented by the 98\textsuperscript{th} percentile, improves from an R\textsuperscript{2} of 75\% to an R\textsuperscript{2} of 94\% (Table 1). Likewise for precipitation, mean daily precipitation improves from an R\textsuperscript{2} of 78\% with the original data to an R\textsuperscript{2} of 85\% after applying ClimaGAN (Fig. 3; Table 2). We also find that extreme precipitation, represented by the 98\textsuperscript{th} percentile, improves from an R\textsuperscript{2} of 72\% to an R\textsuperscript{2} of 78\% (Table 2). The weaker performance of ClimaGAN on distributional skew suggests improvements can be made in capturing aspects of extremes, with one potential cause we are exploring further being the initial normalization steps applied to the data inputs. 

While these performance enhancements from ClimaGAN are promising, we are next curious how they compare against enhancements from a benchmark product, NASA’s NEX-GDDP bias corrected dataset. 

Benchmarking ClimaGAN performance against NASA's product, the first key qualitative distinction is that ClimaGAN outperforms NASA’s product in capturing local spatial variability (Fig. 3). This is because ClimaGAN implements SR to twice the spatial resolution (0.125°) of NASA’s product (0.25°). Visually, the ClimaGAN-enhanced CMIP6 conditions better match observations compared with NASA's product for temperature (Fig. 3). For precipitation, NASA's product appears to better match observations in some areas, in part because it has less spatial variability than ClimaGAN (Fig. 3).

Benchmarking performance quantitatively, we find that ClimaGAN leads to comparable or improved levels of bias correction as NASA's product. For temperature, ClimaGAN outperforms NASA's product on 3 of 4 metrics considered, failing to improve the distributional skew metric (Table 1). Particularly promising is that ClimaGAN yields an R\textsuperscript{2} of 94\% for extreme 98\textsuperscript{th} percentile temperature, compared with 86\% for NASA's product (Table 1). For precipitation, NASA's product outererforms ClimaGAN on all 4 metrics, but the diffrences in performance are small, with R\textsuperscript{2} differences ranging from only 1-3\% (Table 2).

\section{Conclusion}

Here we propose a framework for bias correcting and super-resolving daily climate model inputs to enable more accurate and high spatial resolution simulations of present day and future risk. The framework has several key advantages compared to other commonly employed approaches. First, it allows for superior levels of data-driven spatial resolution enhancement using super-resolution techniques. Second, it jointly bias corrects climate variables, allowing the model to learn from the multivariate relationship between climate variables and more accurately represent multivariate hazards like drought. Third, it flexibly incorporates additional geoscience variables like elevation to inform bias-correction. 

We find that ClimaGAN yields comparable or improved levels of bias-correction at twice the spatial resolution (14km) of NASA’s leading product (25km). This is exciting in part because there are numerous modifications to the network possible that could improve performance (Section 7), while NASA’s product can only improve as the quality of observational data improves. Moreover, NASA typically releases and updates their product once every several years, while ClimaGAN can be regularly updated with the latest sources of data. We expect this ability to update ClimaGAN with the latest observational data as it comes online as well as learn relationships in data-rich regions will help improve bias correction and SR in historically data-poor regions, such as in many developing countries (Section 7). 

Validation results for ClimaGAN suggest substantial potential for high resolution, enhanced accuracy projections of climate risk. Improvements in spatial resolution are critical to capturing local, asset-level effects of climate hazards. We found bias-correction improvements to the raw input data across metrics (Tables 1 and 2), and we highlight that the improvements for extreme precipitation and extreme temperature will enable more accurate projections of hazards like heatwaves and inland flooding. The higher fidelity simulations of present and forward-looking climate variables made possible by applying ClimaGAN can enable more local, accurate models of climate hazards, supporting climate scientists and a broad range of stakeholders alike.

\section{Future Directions}
We see several avenues for expanding and improving the ClimaGAN modeling approach. First, we intend to incorporate additional global regions and CMIP6 models. Second, because ClimaGAN can flexibly integrate additional input channels, we can include variables like humidity, pressure, and wind to not only bias correct those variables but also improve accuracy on the temperature and precipitation variables. Architecture modifications to account for the sparsity of precipitation may further improve results, including distributional skew \cite{Pathak2022}. Last, while we focus here on single image bias correction and SR, we see opportunities for improved performance on day-to-day variability by using multi-temporal images, which has been applied to satellite imagery in the past \cite{deudon2020highres} but never, to our knowledge, to climate model maps.

\bibliographystyle{unsrt}

\clearpage
\end{document}